\documentclass[runningheads]{llncs}
\usepackage[T1]{fontenc}
\usepackage{graphicx}
\usepackage{amsmath}
\usepackage{array}
\usepackage{multirow}
\usepackage{pgfplots}
\usepackage{standalone}
\usepackage{caption}
\usepackage{subcaption}
\usepackage{array}
\usepackage{makecell}
\usepackage{amssymb}
\usepackage{xcolor}
\usepackage{soul}
\usepackage{bbm}

\pgfplotsset{compat=1.8}

\definecolor{darkgreen}{rgb}{0.0, 0.5, 0.0}

\begin{document}
\title{Characterizing the Interpretability of Attention Maps in Digital Pathology}

%


\author{
Tomé Albuquerque\orcidID{0000-0003-3246-7206} \and 
Anil Yüce\orcidID{0000-0003-2688-1873} \and 
Markus D. Herrmann\orcidID{0000-0002-7257-9205} \and 
Alvaro Gomariz\orcidID{0000-0002-6172-5190}}
\authorrunning{T. Albuquerque et al.}
%
\institute{F Hoffmann-La Roche AG, Basel, Switzerland}

%


\maketitle 

\begin{abstract}
Interpreting machine learning model decisions is crucial for high-risk applications like healthcare. In digital pathology, large whole slide images (WSIs) are decomposed into smaller tiles and tile-derived features are processed by attention-based multiple instance learning (ABMIL) models to predict WSI-level labels. These networks generate tile-specific attention weights, which can be visualized as attention maps for interpretability. However, a standardized evaluation framework for these maps is lacking, questioning their reliability and ability to detect spurious correlations that can mislead models. We herein propose a framework to assess the ability of attention networks to attend to relevant features in digital pathology by creating artificial model confounders and using dedicated interpretability metrics. Models are trained and evaluated on data with tile modifications correlated with WSI labels, enabling the analysis of model sensitivity to artificial confounders and the accuracy of attention maps in highlighting them. Confounders are introduced either through synthetic tile modifications or through tile ablations based on their specific image-based features, with the latter being used to assess more clinically relevant scenarios. We also analyze the impact of varying confounder quantities at both the tile and WSI levels. Our results show that ABMIL models perform as desired within our framework. While attention maps generally highlight relevant regions, their robustness is affected by the type and number of confounders. Our versatile framework has the potential to be used in the evaluation of various methods and the exploration of image-based features driving model predictions, which could aid in biomarker discovery.

\end{abstract}

\section{Introduction}
The integration of digital pathology (DP), particularly with images of Hematoxylin and Eosin (H\&E) stained tissue sections, in clinical practice opens the possibility to advance diagnostics through artificial intelligence (AI).
Machine learning (ML) models could assist diagnostic tasks by discovering input-output correlations in training data. However, these models usually rely on correlations rather than causal connections~\cite{Baaj2024}. Consequently, ML predictions may be based on spurious correlations, including confounding artifacts such as pen marks on an image~\cite{Plass2023}. Such confounders can  inaccurately inflate prediction performance, posing risks when human practitioners depend on these predictions. Indeed, the lack of explainability of black-box models is a significant barrier to implementing AI in high-stakes fields, namely in healthcare~\cite{Miao2021}. Therefore, tools to understand and trace ML decision processes are essential to ensure reliable AI support in medical decisions.
The rapid advancement of explainable AI has led to numerous techniques enhancing the transparency of black-box ML systems~\cite{Hassija2023}. 
However, the intricacies of DP datasets call for the use of dedicated methodologies. 

\emph{Whole slide images (WSI)} in DP are very large and usually cannot be processed by typical deep learning classification models due to computational constraints. The multiple instance learning (MIL) framework has been adopted in the community, where WSIs are decomposed into smaller \emph{tiles} that inherit the WSI's label during training~\cite{maron1997framework}. The popular attention-based MIL (ABMIL)~\cite{ilse2018attention} network employs a global attention mechanism to learn weights accounting for the contribution of each tile representation to the final WSI prediction. Tiles in the WSI are often visualized by their weights values as attention maps, enabling the interpretation of which regions are prioritized by the ABMIL model prediction process~\cite{bodria2021benchmarking}.

The described interpretability methods make ML predictions more transparent and may reveal new pathological insights. However, the proficiency of attention maps in identifying misleading correlations, also known as spurious correlations, has been criticized~\cite{Julius2018,ghassemi2021false}.
In the context of X-ray chest images, a framework has been proposed~\cite{Sun2023} to assess the identification of spurious correlations in classification models by systematically introducing confounders in the training and evaluation data.  

\begin{figure}[ht]
    \centering
    \includegraphics[width=0.85\textwidth]{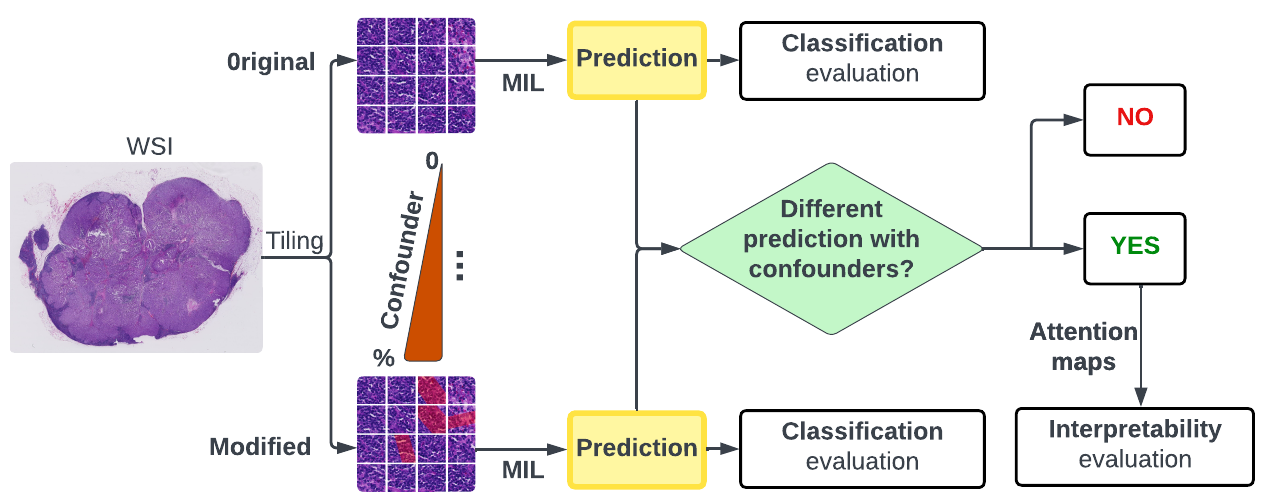}
    \caption{Overview of our framework for the evaluation of attention maps in DP.}
    \label{fig:diagram}
\end{figure}

In the DP ABMIL setting, only some of the tiles may contain the information necessary for model predictions, hence requiring a different experimental design and evaluation metrics for the assessment of attention maps employed for interpretability.
We herein build on the work of Sun et al.~\cite{Sun2023} by providing a framework suitable for ABMIL with DP WSIs, including new experimental designs and metrics that emphasize the role of tiles in model predictions. 
Our framework, illustrated in Fig.\ \ref{fig:diagram}, includes the study of tile- and WSI-based confounders, which are first validated with synthetic confounders involving image modifications. We then explore the more clinically relevant scenario of enriching for confounders via sampling tiles based on image-based features. 

\section{Methods}
\subsection{Attention Maps in Digital Pathology}
We adopt the ABMIL~\cite{ilse2018attention} framework described above for binary classification. 
A WSI $s_i, \, i \in \{1, \ldots, S\}$ is decomposed into non-overlapping tiles $t_i^j \in \mathbb{R}^{n \times n}$, with $n$ the tile size and $S$ the number of WSIs. 
Tiles are processed by a feature extraction model ${G: \mathbb{R}^{n \times n} \rightarrow \mathbb{R}^k}$, yielding embedding vectors of length $k$.
We define the ABMIL model as ${F: \mathbb{R}^{T \times k} \rightarrow \{0,1\}}$, which employs an attention mechanism on the $T$ embeddings to predict the binary WSI label, with ground truth $y_{s_i}$.
Attention maps are generated by visualizing the resulting attention weights ${a_i^j \in \mathbb{R}}$ corresponding to each of the tiles $t_i^j$.
These maps help identify which regions contribute most to the WSI-level prediction, enhancing interpretability and aiding in the detection of spurious correlations or areas of interest.

\subsection{Framework for the Interpretability of Attention Maps}
As illustrated in Fig.\ \ref{fig:diagram}, our framework applies controlled modifications to the tiles in WSIs of a specific class. These modifications act as spurious correlations that aid the classification model. A confounded model may change its prediction when such modification is also used in a test image. When this occurs, we can assess whether the attention map highlights the modified tiles, hence enabling our evaluation. 

Let $M_p \sim \textit{Bernoulli}(p)$ be a Bernoulli random function. 
We define $M_p$ separately for tiles ($M_p^t$) and WSIs ($M_p^s$).
This means that a modified WSI $\hat{s}^i$ is produced when $M_p^s(s_i)=1$, which occurs with probability $p$. 
As the goal of $M_p$ is to introduce confounders, it is only applied to WSIs with positive label. Consequently, $M_p^s(s_i)$ is always 0 for WSIs with negative label ($y_{s_i}=0$).
Only modified WSIs $\hat{s}^i$ can have modified tiles $\hat{t}_i^j$, as defined by $M_p(t_i^j)$. Thus, within a modified WSI, only some tiles are actually modified.
Two different modification designs are employed in our experimental setting:

$\bullet$ \textit{Tile-based:} Different percentages of tiles are modified for all WSIs. Specifically, $M_p^s$ is set to 1, and a range of $M_p^t$ values is assessed. This strategy enables the examination of how the amount of systematic confounders (i.e., present in all WSIs) influences the model.

$\bullet$ \textit{WSI-based:} We study the effect of sporadic confounders that only affect a varying number of WSIs, whose tiles are affected with the same probability, i.e.\ a range of $M_p^s$ is assessed while $M_p^t$ is set to 0.5. This value of $M_p^t$ is heuristically defined based on the tile-based observations.

\subsection{Evaluation of Artificially Confounded Datasets}
ABMIL models $F$ are trained for different training sets created with different modifications $M_p$. Two types of evaluations are performed for assessing the predictions and interpretability separately:

\vspace{1ex}\noindent{\bf Model predictions: }
The performance of $F$ is measured with the Area Under the receiver operator characteristics Curve (AUC) in the binary classification task. A test set is employed with the same $M_p$ process as the training set of that model. The goal is to assess whether the performance increases with $p$, which would confirm that the modification is indeed acting as a confounder and aiding the model in an artificial way, hence enabling the interpretability evaluation.

\vspace{1ex}\noindent{\bf Interpretability: }
Each model $F$ is evaluated on each WSI of the test set, once with and once without modification ($F(\hat{s}_i)$ and $F(s_i)$ respectively). 
The goal is to assess attention differences in the subset of WSIs where the modification alters the prediction, defined as $\bar{S} = \{s_i \mid F(\hat{s}_i) \neq F(s_i)\}$.
These differences are measured as a confusion matrix, where modified tiles $\hat{t_i^j}$ are considered positives, and unmodified tiles $t_i^j$ are considered negatives.
The predicted attention $a_i^j$ is considered a positive prediction when it is in the top 20\% of attention tiles, and negative otherwise. We use the function $F_\text{att}^\text{top}$ to denote when a tile is among the top 20\% tiles ($F_\text{att}^\text{top}(t_i^j)=1$) or not ($F_\text{att}^\text{top}(t_i^j)=0$).
Two metrics are used in our evaluation:

$\bullet$ \emph{Confounder Robustness (CR): } We propose this metric to quantify the ability of an ABMIL model to attribute its decision to a confounder, specifically as the ratio of WSIs for which the attention to modified tiles is better than random guess. 
With the confounder matrix as defined above, it follows that $\textit{Prevalence}=\left(1/T\right)\sum_{t_i^j \in s_i}M(t_i^j)$ and 
\begin{equation*}
    \textit{Precision} = \frac{\sum_{t_i \in \bar{S}}F_\text{att}^\text{top}(t_i^j) M(t_i^j)}{\sum_{t_i \in \bar{S}}F_\text{att}^\text{top}(t_i^j)}
\end{equation*}
Considering that by definition a random assignment would, in average, match the precision to the prevalence, we define the CR metric as:
\begin{equation*}
    \textit{CR} = \frac{1}{\bar{S}}\sum_{s \in \bar{S}} \mathbbm{1} \left( \textit{Precision} > \textit{Prevalence} \right)
\end{equation*}

$\bullet$ \emph{Normalized Cross Correlation (NCC):} This metric measures the sensitivity to changes in attention when applying a modification. 
\begin{equation*}
    \textit{NCC} = \frac{1}{\bar{S}} \sum_{s_i \in \bar{S}} \frac{\sum_{t_i^j \in s_i} \left( a_i^j - \overline{a_i} \right) \left( \hat{a}_i^j - \overline{\hat{a}_i} \right) } {\sum_{t_i^j \in s_i} \sqrt{\left( a_i^j - \overline{a_i} \right)^2 \left( \hat{a}_i^j - \overline{\hat{a}_i} \right)^2}}
\end{equation*}
where the mean of the attention across tiles in $s_i$ is denoted as  $\overline{a_i}$ and $\overline{\hat{a}_i}$ for its original and modified version respectively. 

In this framework, $\text{NCC}$$=$$1$ and $\text{CR}$$=$$0$ when $p$$=$$0$, as there are no modifications and the model produces the same predictions. For low $p$, the model may not be confounded yet, hence not paying attention to the modifications and yielding high NCC and low CR values. When $p$$=$$1$, the model should be completely confounded and base its predictions only on the existence of confounders in the test data, which, in an ideal scenario, would lead to $\text{NCC}$$=$$0$ and $\text{CR}$$=$$1$.

\section{Experiments and Results}
\noindent{\bf Dataset:}
We employ CAMELYON16~\cite{CAMELYON16}, a publicly available dataset formed by 399 WSIs of sentinel
lymph node tissue sections used for binary classification of WSIs into lymph node tissue with (60\%) or without (40\%) cancer metastasis. 
We use all tissue tiles and follow the train-test split proposed for the original challenge, with $20\%$ of the training data employed as validation set, which is employed to select the best training epoch based on the loss value.

\vspace{1ex}\noindent{\bf Implementation:}
WSIs are decomposed into patches of size $n$$=$$256$ pixels, which are processed by a ResNet50 pre-trained on DP samples as described in~\cite{abbasi2021molecular,gildenblat2023deep,bredell2023aggregation}.
The resulting embeddings are aggregated using ABMIL with bags of of 1024 tiles, 32 bags per batch, a maximum of $300$ epochs, feature layers with $[1024, 1024, 512, 128, 64, 32]$ nodes, dropout rate of $0.1$, binary crossentropy loss with label smoothing, and learning rate $0.0001$.

\subsection{Synthetic Modifications}
\begin{figure}[ht]
    \centering
    \begin{tabular}{c|ccc}
             Original & Clever Hans & Blur & Pen mark\\
             \includegraphics[width=2.2cm]{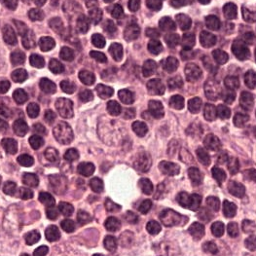} &       \includegraphics[width=2.2cm]{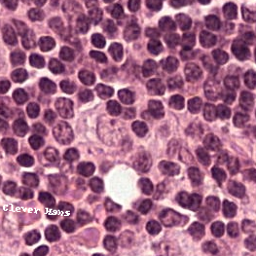} &       \includegraphics[width=2.2cm]{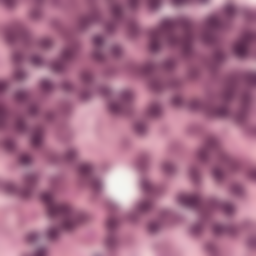} &       \includegraphics[width=2.2cm]{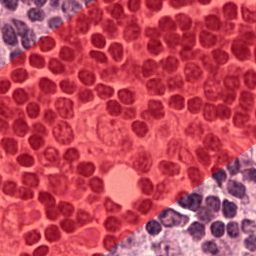}\\
    \end{tabular}
    \caption{Examples of synthetic tile modifications employed.}
    \label{fig:dataset}
\end{figure}

We employ transformations that change the appearance of individual tiles as indicated in Fig.\ \ref{fig:dataset}, which allow for controlled experiments to assess the impact of these modifications both on the classification and interpretability performance.
Note that the modifications are only added to the metastatic tissue class to act as model confounders.
The \emph{Clever Hans} modification is created by inserting text ("Clever Hans") into the tile at a random position, applying random rotation, and overlaying it onto the original image using alpha compositing. 
The \emph{blur} artifact is achieved using a Gaussian filter with a standard deviation of $4$ pixels. 
The \emph{pen mark} modification is generated with two random points, between which a red line is drawn and overlaid with alpha compositing.

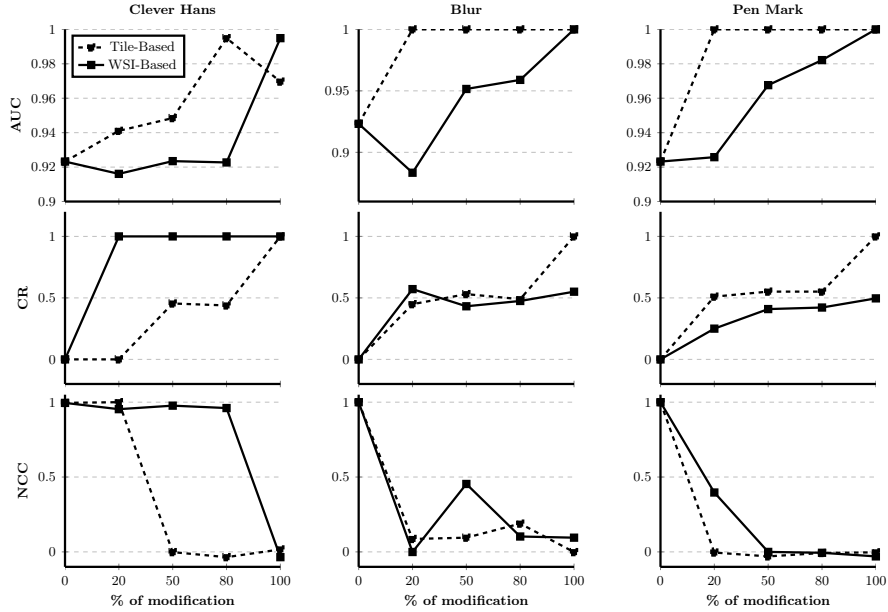
\begin{figure}[ht]
    \centering
    \begin{subfigure}{0.32\textwidth}
        \centering
        \resizebox{\linewidth}{!}{\pgfplotsset{compat=1.8,width=5cm, height=4cm, scale only axis}
\begin{tikzpicture}

    \begin{axis}[name=plot1,
        ymin=0.9, ymax=1, 
        xtick={1, 2, 3, 4, 5},
        xticklabels={}, 
        ymajorgrids=true,
        legend pos=north west,
        ylabel={\textbf{AUC}},
        grid style=dashed,
        axis y line=left,
        axis x line=bottom,
        axis line style={-}, 
        ylabel near ticks,
        xlabel near ticks,
        line width=1.5pt,
        title=\textbf{Clever Hans}
    ]

    \addplot[mark=square*, dashed, black] coordinates {(1,0.9232142857142858)(2,0.9410714285714286)(3,0.948469387755102)(4,0.9948979591836735)(5,0.969642857142857)};
    \addlegendentry{Tile-Based}

    \addplot[mark=square*, black] coordinates {(1,0.9232142857142858)(2,0.9160714285714285)(3,0.923469387755102)(4,0.922704081632653)(5,0.9948979591836735)};
    \addlegendentry{WSI-Based}
    
    \end{axis}

    \begin{axis}[name=plot2, at={(plot1.below south west)}, anchor=above north west,
                 yshift=3pt, ymajorgrids=true,
                 xtick={1, 2, 3, 4, 5}, xticklabels={},
                 grid style=dashed, ylabel={\textbf{CR}}, ymin=-0.2, ymax=1.2,
                 axis y line=left,
                 axis x line=bottom,
                 axis line style={-}, 
                 ylabel near ticks,
                 xlabel near ticks,
                 line width=1.5pt,
    ]

    \addplot[mark=square*, dashed, black] coordinates {(1,0.0)(2,0.0)(3,0.45454545454545453)(4,0.43859649122807015)(5,1.0)};

    \addplot[mark=square*, black] coordinates {(1,0.0)(2,1.0)(3,1.0)(4,1.0)(5,1.0)};

    \end{axis}

    \begin{axis}[name=plot3, at={(plot2.below south west)}, anchor=above north west,
                 yshift=3pt, ymajorgrids=true, grid style=dashed,
                 xtick={1, 2, 3, 4, 5}, xticklabels={0, 20, 50, 80, 100},
                 xlabel={\textbf{\% of modification}}, ylabel={\textbf{NCC}}, ymin=-0.1, ymax=1.05,
                 axis y line=left,
                 axis x line=bottom,
                 axis line style={-}, 
                 ylabel near ticks,
                 xlabel near ticks,
                 line width=1.5pt,
    ]

    \addplot[mark=square*, dashed, black] coordinates {(1,0.99505913)(2,1.0)(3,-0.0016565386)(4,-0.035271168)(5,0.016430186)};

    \addplot[mark=square*, black] coordinates {(1,0.9951741)(2,0.95329076)(3,0.9767721)(4,0.9607997)(5,-0.035271168)};

    \end{axis}
 
\end{tikzpicture}}
        \label{fig:clever}
    \end{subfigure}%
    \begin{subfigure}{0.32\textwidth}
        \centering
        \resizebox{\linewidth}{!}{\pgfplotsset{compat=1.8,width=5cm, height=4cm, scale only axis}
\begin{tikzpicture}

    \begin{axis}[name=plot1,
        ymin=0.86, ymax=1, 
        xtick={1, 2, 3, 4, 5},
        xticklabels={}, 
        ymajorgrids=true,
        legend pos=north west,
        ylabel={\textcolor{white}{\textbf{AUC}}},
        grid style=dashed,
        axis y line=left,
        axis x line=bottom,
        axis line style={-}, 
        ylabel near ticks,
        xlabel near ticks,
        line width=1.5pt,
        title=\textbf{Blur}
    ]

    \addplot[mark=square*, dashed, black] coordinates {(1,0.9232142857142858)(2,1.0)(3,1.0)(4,1.0)(5,1.0)};
    \addplot[mark=square*, black] coordinates {(1,0.9232142857142858)(2,0.8834183673469388)(3,0.951530612244898)(4,0.9589285714285715)(5,1.0)};

    \end{axis}

    \begin{axis}[name=plot2, at={(plot1.below south west)}, anchor=above north west,
                 yshift=3pt, ymajorgrids=true,
                 xtick={1, 2, 3, 4, 5}, xticklabels={},
                 grid style=dashed, ylabel={\textcolor{white}{\textbf{CS}}}, ymin=-0.2, ymax=1.2,
                 axis y line=left,
                 axis x line=bottom,
                 axis line style={-}, 
                 ylabel near ticks,
                 xlabel near ticks,
                 line width=1.5pt,
    ]

    \addplot[mark=square*, dashed, black] coordinates {(1,0.0)(2,0.4489795918367347)(3,0.5306122448979592)(4,0.4897959183673469)(5,1.0)};

    \addplot[mark=square*, black] coordinates {(1,0.0)(2,0.5714285714285714)(3,0.4318181818181818)(4,0.475)(5,0.5503875968992248)};

    \end{axis}

    \begin{axis}[name=plot3, at={(plot2.below south west)}, anchor=above north west,
                 yshift=3pt, ymajorgrids=true, grid style=dashed,
                 xtick={1, 2, 3, 4, 5}, xticklabels={0, 20, 50, 80, 100},
                 xlabel={\textbf{\% of modification}}, ylabel={\textcolor{white}{\textbf{NCC}}}, ymin=-0.1, ymax=1.05,
                 axis y line=left,
                 axis x line=bottom,
                 axis line style={-}, 
                 ylabel near ticks,
                 xlabel near ticks,
                 line width=1.5pt,
    ]

    \addplot[mark=square*, dashed, black] coordinates {(1,1)(2,0.086331196)(3,0.09488767)(4,0.18873888)(5,-0.001512315)};


    \addplot[mark=square*, black] coordinates {(1,1)(2,-0.0007739085)(3,0.4536737)(4,0.102708355)(5,0.09488767)};
    \end{axis}
 
\end{tikzpicture}}
        \label{fig:blur}
    \end{subfigure}
    \begin{subfigure}{0.32\textwidth}
        \centering
        \resizebox{\linewidth}{!}{\pgfplotsset{compat=1.8,width=5cm, height=4cm, scale only axis}
\begin{tikzpicture}

    \begin{axis}[name=plot1,
        ymin=0.9, ymax=1, 
        xtick={1, 2, 3, 4, 5},
        xticklabels={}, 
        ymajorgrids=true,
        legend pos=north west,
        ylabel={\textcolor{white}{\textbf{AUC}}},
        grid style=dashed,
        axis y line=left,
        axis x line=bottom,
        axis line style={-}, 
        ylabel near ticks,
        xlabel near ticks,
        line width=1.5pt,
        title=\textbf{Pen Mark}
    ]

    \addplot[mark=square*, dashed, black] coordinates {(1,0.9232142857142858)(2,1.0)(3,1.0)(4,1.0)(5,1.0)};
    
    \addplot[mark=square*, black] coordinates {(1,0.9232142857142858)(2,0.925765306122449)(3,0.9676020408163265)(4,0.9821428571428572)(5,1.0)};

    \end{axis}

    \begin{axis}[name=plot2, at={(plot1.below south west)}, anchor=above north west,
                 yshift=3pt, ymajorgrids=true,
                 xtick={1, 2, 3, 4, 5}, xticklabels={},
                 grid style=dashed, ylabel={\textcolor{white}{\textbf{CS}}}, ymin=-0.2, ymax=1.2,
                 axis y line=left,
                 axis x line=bottom,
                 axis line style={-}, 
                 ylabel near ticks,
                 xlabel near ticks,
                 line width=1.5pt,
    ]

    \addplot[mark=square*, dashed, black] coordinates {(1,0.0)(2,0.5102040816326531)(3,0.5510204081632653)(4,0.5510204081632653)(5,1.0)};

    \addplot[mark=square*, black] coordinates {(1,0.0)(2,0.25)(3,0.4090909090909091)(4,0.42168674698795183)(5,0.49612403100775193)};

    \end{axis}

    \begin{axis}[name=plot3, at={(plot2.below south west)}, anchor=above north west,
                 yshift=3pt, ymajorgrids=true, grid style=dashed,
                 xtick={1, 2, 3, 4, 5}, xticklabels={0, 20, 50, 80, 100},
                 xlabel={\textbf{\% of modification}}, ylabel={\textcolor{white}{\textbf{NCC}}}, ymin=-0.1, ymax=1.05,
                 axis y line=left,
                 axis x line=bottom,
                 axis line style={-}, 
                 ylabel near ticks,
                 xlabel near ticks,
                 line width=1.5pt,
    ]

    \addplot[mark=square*, dashed, black] coordinates {(1,1)(2,-0.004292645)(3,-0.030148543)(4,-0.007330433)(5,-0.0029061974)};

    \addplot[mark=square*, black] coordinates {(1,1)(2,0.39662558)(3,-0.00013304439)(4,-0.006063666)(5,-0.030148543)};

    \end{axis}
 
\end{tikzpicture}}
        \label{fig:penmark}
    \end{subfigure}
    \caption{Classification (top) and explainability performance results (middle and bottom) for  synthetic experiments.}
    \label{fig:synthetic}
\end{figure}

\vspace{1ex}\noindent{\bf Classification Performance:}
The AUC in Fig.\ \ref{fig:synthetic} shows that, with a few minor exceptions, the ABMIL model gradually obtains better classification results as the amount of confounders increases, both for the tile- and WSI-based experiments, and for all three types of modifications.
This performance improvement confirms that the synthetic modifications act as a confounder for the model. These results highlight the model's susceptibility to spurious correlation, hence enabling an accurate evaluation of the interpretability of the attention maps. 
Interestingly, the increase in performance is generally lower for the WSI-based approach. 
This suggests that a lower amount of systematic (tile-based) confounders can mislead the model more easily than a higher amount of sporadic (WSI-based) confounders.
The varying influence of each type of modification underscores the different effects that various confounders can have.

\vspace{1ex}\noindent{\bf Interpretability Performance:}
The explainability metrics in Fig.\ \ref{fig:synthetic} confirm that, as the amount of confounders increases, the attention maps tend to focus more on the modified tiles. This effect is reflected by the fact that, as the \% of modification $p$ increases, the NCC decreases and the CS increases, which would otherwise be close to 1 and 0 respectively for any degree of modifications.
When the WSI-based \% of modifications is low, the  CR increases substantially more for Clever Hans than for the other transformations. This observation highlights the dependence of model's sensitivity to different confounders, which in our setup is lower for Clever Hans.
The low value of NCC for the blur experiments is also relevant. 
We hypothesize this is due to the difficulty in extracting pertinent information from blurred tiles, which creates higher variations in attention weights.

\subsection{Feature-based Sampling Strategy}
While the synthetic confounders presented above allow for controlled experiments in our framework, the results may not be representative of real-world cases with more subtle confounders. 
We herein present a proof of concept using tile ablations to dilute an existing signal, namely by leveraging the known influence of real WSI-derived features on the model prediction.
The size of nuclei of breast cancer cells is known to be more variable than the size of nuclei of normal lymphocytes in the lymph nodes~\cite{John1991}. Hence a higher variance in nuclear size can be leveraged as a proxy for the presence of breast cancer metastases.
We employ the \emph{Standard Deviation of the Average Nuclear Area} (SDANA), which is calculated by first taking the average area of the cell nuclei for each tile and then taking the standard deviation of these values across the entire WSI.
SDANA is herein studied as a feature that the ABMIL model $F$ leverages for its predictions.

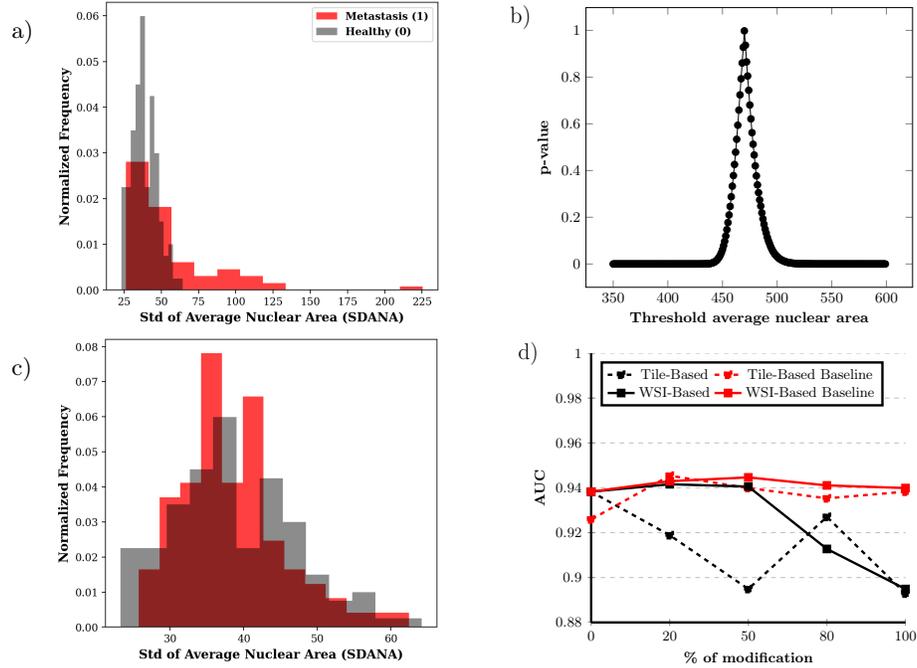
\begin{figure}[ht]
    \centering
    \begin{subfigure}{0.45\textwidth}
        \centering
        \vspace{0.1cm}
        \begin{tikzpicture}
        \node[inner sep=0pt] (russell) at (0,0){    \includegraphics[width=0.95\linewidth]{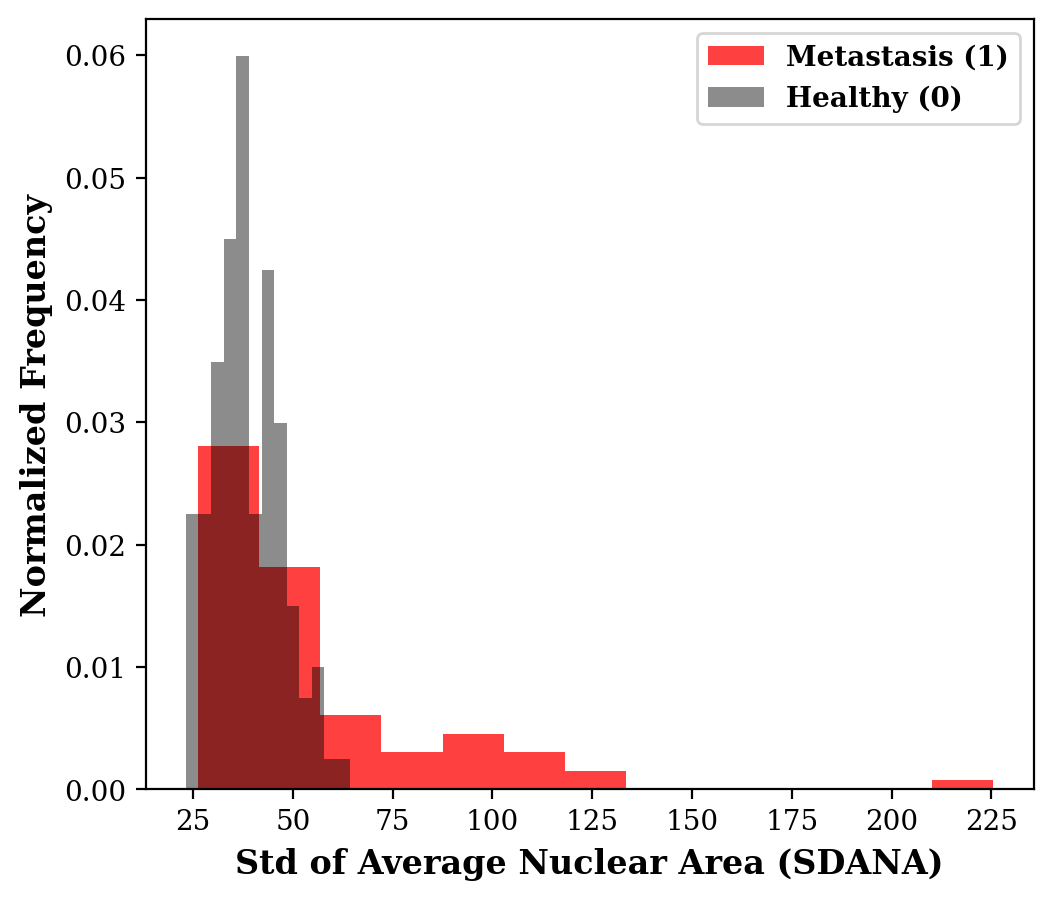}};
        \node at (-3,1.75) {a)};
        \end{tikzpicture}
        \captionsetup{labelformat=empty}
        \caption[]{}
        \label{fig:oridist}
    \end{subfigure}%
    \hspace{1cm}
    \begin{subfigure}{0.45\textwidth}
        \centering
        \resizebox{\linewidth}{!}{%
        \pgfplotsset{width=7cm, height=6cm, scale only axis}
\begin{tikzpicture}
\begin{axis}[
  xlabel={\textbf{Threshold average nuclear area}}, 
  ylabel={\textbf{p-value}}, 
]
\addplot [color = black ,mark=*] table [x=x, y=y, col sep=comma] {figures/area_mean_pvalue_th.csv};

\end{axis}
\node at (-1.4,5.8) {\Large b)};
\end{tikzpicture}}
        \captionsetup{labelformat=empty}
        \caption[]{}
        \label{fig:pvalueth}
    \end{subfigure}
    \hfill
    \vspace{-0.6cm} 
    \begin{subfigure}{0.45\textwidth}
        \centering
        \begin{tikzpicture}
        \node[inner sep=0pt] (russell) at (0,0){\includegraphics[width=0.95\linewidth]{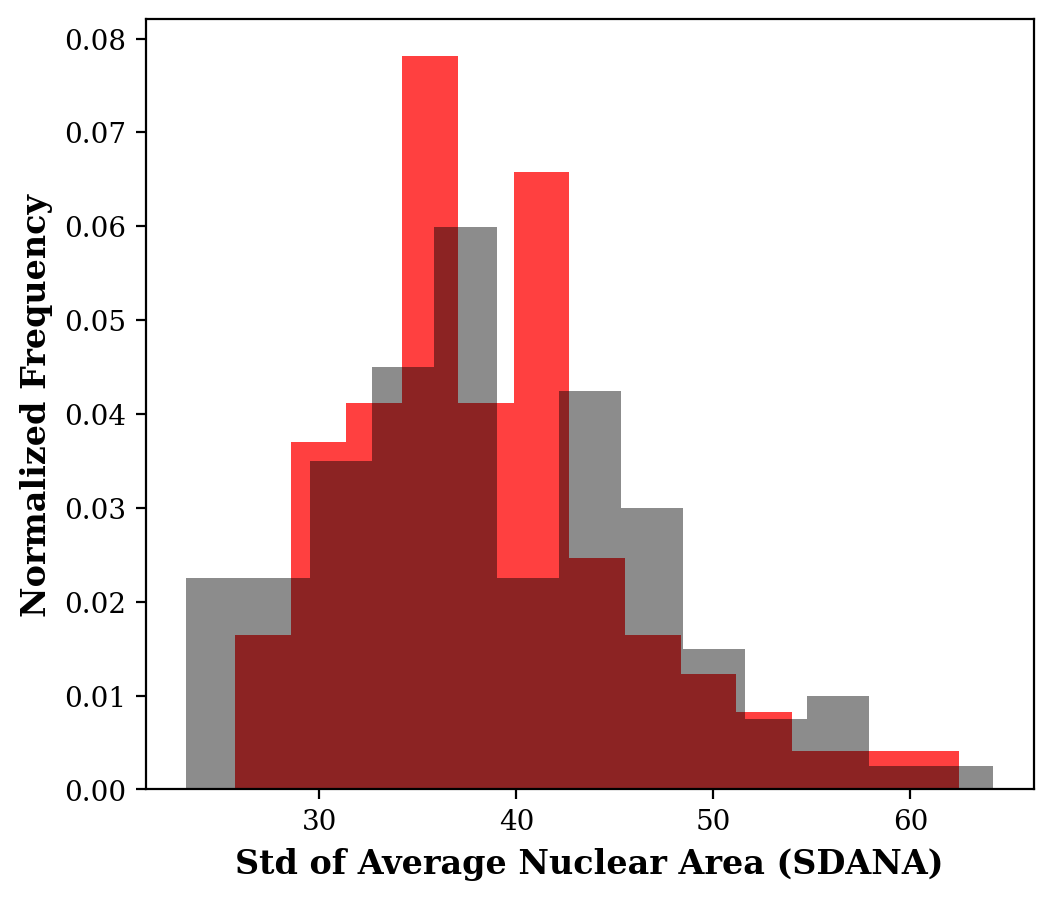}};
        \node at (-3,1.75) {c)};
        \end{tikzpicture}
        \captionsetup{labelformat=empty}
        \caption[]{}
        \label{fig:thdist}
    \end{subfigure}%
    \vspace{-0.7cm}
    \hfill
    \begin{subfigure}{0.45\textwidth}
        \centering
        \resizebox{\linewidth}{!}{\pgfplotsset{compat=1.8,width=7cm, height=6cm, scale only axis}
\begin{tikzpicture}

    \begin{axis}[name=plot1,
        ymin=0.88, ymax=1, 
        xtick={1, 2, 3, 4, 5}, xticklabels={0, 20, 50, 80, 100}, xlabel={\textbf{\% of modification}},
        legend pos=north west,
        legend columns=2,
        ymajorgrids=true,
        ylabel={\textbf{AUC}},
        grid style=dashed,
        axis y line=left,
        axis x line=bottom,
        axis line style={-},
        ylabel near ticks,
        xlabel near ticks,
        line width=1.5pt,
    ]

    \addplot[mark=square*,dashed, black] coordinates {(5,0.8928571428571428)(4,0.9270408163265307)(3,0.8948979591836734)(2,0.9188775510204082)(1,0.938265306122449)};
    \addlegendentry{Tile-Based}

    \addplot[mark=square*,dashed, red] coordinates {(5,0.938265)(4,0.935306)(3,0.939898)(2,0.945408)(1,0.925918)};
    \addlegendentry{Tile-Based Baseline}

    
    \addplot[mark=square*, black] coordinates {(1,0.938265306122449)(2,0.9415816326530613)(3,0.9405612244897958)(4,0.9127551020408163)(5,0.8948979591836734)};
    \addlegendentry{WSI-Based}

    \addplot[mark=square*, red] coordinates {(1,0.938265)(2,0.942959)(3,0.944643)(4,0.941071)(5,0.939898)};
    \addlegendentry{WSI-Based Baseline}

    \end{axis}




\node at (-1.4,6) {\Large d)};
\end{tikzpicture}}
        \captionsetup{labelformat=empty}
        \caption[]{}
        \label{fig:qpfsres}
    \end{subfigure}
    \caption{Feature-based sampling strategy. (a) Original distribution. (b) SDANA class separation. (c) Distribution after ablation. (d) Classification results.}
    \label{fig:qpfs}
\end{figure}

In this setting, the prevalence of specific tiles is systematically reduced with the goal of obtaining a similar SDANA in WSIs from either class, so that the model cannot leverage this feature. 
The original training set has the distribution of SDANA per WSI shown in Fig.\ \ref{fig:oridist}, with a clear separation between metastatic and healthy tissue WSIs. 
We analyze how to reduce this separation by selectively removing tiles, only for metastatic tissue WSIs, with an average cell area above a specific threshold. 
Fig.\ \ref{fig:pvalueth} shows the resulting p-value, as measured by a t-test, of the separation between both distributions (SDANA for healthy and ablated WSIs) for different thresholds. The p-value is maximized at threshold 470, which is used hereon. 
By eliminating the tiles in the metastatic tissue WSIs with a mean area of cells above $470$, we bring the class distributions for SDANA as close as possible, as shown in Fig.\ \ref{fig:thdist}, and hence prevent the ABMIL model from using SDANA as a discriminative feature.

In this setting, the modification function $M_p$ is not random. Instead, $p$ denotes the ratio of tiles with cell area above the selected threshold of 470 to be removed. The removed tiles are chosen based on their cell area from highest to lowest.
This setting only accounts for classification performance, since our interpretability metrics require a consistent number of modified tiles; however, in this setting tiles are removed. 
Baseline results with a random ablation are included for comparison. 
This baseline is based on an equivalent ablation, where the same number of tiles are removed as in the feature-based experiment, with the difference that they are selected randomly across the WSI. 
The ablation for this random baseline is repeated with 5 replicates using different random tiles, for which the average AUC is employed.

The results in Fig.\ \ref{fig:qpfsres} show that, as hypothesized, the AUC decreases as the amount of removed tiles increases, which brings the SDANA distributions closer across both classes. 
This is a realistic scenario with less control on the data, which explains why the trend is not as clear as with the synthetic experiment results. Still, the decrease in performance is clear when compared to the random baseline experiment. 
Hence, this feature-based framework enables the study of the relevance of specific concept features for DP classification tasks, further improving our understanding of the model's decision-making process.

\section{Conclusions}
We herein establish a framework for evaluating the ability of attention maps to highlight regions in a WSI containing patterns that correlate with the WSI-level label. 
Our results based on synthetic tile ablations underscore the value of this framework, as both the classification and interpretability metrics follow the expected trends. 
The results further confirm that the robustness and interpretability of ABMIL models depend on the type and systematic nature of confounders.
Additionally, we propose a feature-based sampling strategy, where real DP features are employed for diluting the signal, i.e.\ by reducing the prevalence of tiles containing patterns known to be relevant for the task. 
This setting demonstrates the value of our framework in real-world scenarios, where the impact of specific DP features can be quantitatively measured. 
Our framework can also be extended to evaluate real-world confounders that may negatively affect ML models, which could be assessed on a dataset where natural confounders have been labeled.
Moreover, different feature extraction models and attention mechanisms could be evaluated in our framework to characterize their behavior in the presence of confounders. 
This versatility enhances the potential application of our framework, paving the way for more robust and interpretable AI solutions in digital pathology.

\newpage
\bibliographystyle{splncs04}
\bibliography{refs}
\end{document}